\documentclass[12pt,letterpaper]{article}
\usepackage{amsmath}
\usepackage{amssymb}
\usepackage{latexsym}
\newcommand{\be}{\begin{equation}}
\newcommand{\ee}{\end{equation}}
\newcommand{\bea}{\begin{eqnarray}}
\newcommand{\eea}{\end{eqnarray}}

\newcommand{\calh}{{\mathcal H}}

\newcommand{\calo}{{\mathcal O}}

\newcommand{\dS}{{\mathrm{dS}}}
\newcommand{\OO}{{\mathrm{OO}}}
\newcommand{\BB}{{\mathrm{BB}}}

\begin{document} 

\title{Why Boltzmann Brains Are Bad}

 \author{Sean M. Carroll\\
 Walter Burke Institute for Theoretical Physics\\
 California Institute of Technology, Pasadena, CA 91125, U.S.A.\\
 seancarroll@gmail.com}
\maketitle

\begin{quotation} \noindent
Some modern cosmological models predict the appearance of Boltzmann Brains: observers who randomly fluctuate
out of a thermal bath rather than naturally evolving from a low-entropy Big Bang. 
A theory in which most observers are of the Boltzmann Brain type is generally thought to be unacceptable, although opinions differ. I argue that such theories are indeed unacceptable: the real 
problem is with fluctuations into observers who are locally identical to ordinary observers, and their existence
cannot be swept under the rug by a choice of probability distributions over observers. The issue is not
that the existence of such observers is ruled out by data, but that the theories that predict them are cognitively
unstable: they cannot simultaneously be true and justifiably believed.
\end{quotation}

\tableofcontents


\newpage

\begin{quote}
\emph{If you want to make an apple pie from scratch, you must first invent the universe.} -- Carl Sagan
\end{quote}

\section{Introduction}

Modern cosmology has put us in a predicament that Ren\'e Descartes had worked to escape from in the seventeenth century.
In a famous thought experiment in his \emph{Meditations on First Philosophy}, Descartes sought a basis for absolutely certain knowledge.
You might think that the floor on which you are standing certainly exists, but Descartes worried that this impression could be mistaken; maybe you are actually dreaming, or you are being misled by a powerful, evil demon.
(Today we might fear that we are a brain in a vat, or living in a computer simulation.)
Descartes eventually decides that there is something he can't consistently doubt: namely, that he is thinking these doubts, and that therefore he must exist.
That is something about which he can be absolutely certain.
The road from his own existence to the reliability of his senses is a convoluted one, going through an ontological argument for the existence of God, about which later commentators have been justifiably skeptical.

The issue is undoubtedly important. 
We use the evidence of our senses as the basis of the scientific process; we collect data, and use that to test and refine models that describe the world. 
If that evidence were completely unreliable, science itself would be impossible.
But even without a Cartesian bedrock of certainty to undergird trust in our senses, it seems reasonable (at least for working scientists, if not always for more fastidious philosophers) to take the reliability of what we see as an assumption.
Absent some strong reason to believe that there actually is an evil demon trying to trick us, it seems natural to put a high credence in the basic accuracy of our manifest picture of the world, and work from there.

The problem arises when we start with our experience of the world, use it to construct our best theory of how the world actually operates, and then realize that this theory itself predicts that our sense data are completely unreliable.

That is the conundrum we are potentially put in by the problem of Boltzmann Brains (BBs) \cite{Dyson:2002pf,Albrecht:2004ke,Carroll:2010zz}. 
In brief, the BB problem arises if our universe (1) lasts forever (or at least an extraordinarily long time, much longer than $10^{10^{66}}$ years), and (2) undergoes random fluctuations that could potentially create conscious observers.
If the rate of fluctuations times the lifetime of the universe is sufficiently large, we would expect a ``typical'' observer to be such a fluctuation, rather than one of the ordinary observers (OOs) that arise through traditional thermodynamic evolution in the wake of a low-entropy Big Bang.
We humans here on Earth have a strong belief that we are OOs, not BBs, so there is apparently something fishy about a cosmological model that predicts that almost all observers are BBs.

This mildly diverting observation becomes more pressing if we notice that the current best-fit model for cosmology -- denoted $\Lambda$CDM, where $\Lambda$ stands for the cosmological constant (vacuum energy) and CDM for ``cold dark matter'' -- is arguably a theory that satisfies both conditions (1) and (2).
(Many eternally-inflating cosmologies potentially do so as well.)
If the vacuum energy remains constant, the universe asymptotically approaches a de~Sitter phase.
Such a spacetime resembles, in certain ways, a box of gas in equilibrium, with a de~Sitter temperature $T_\dS = \sqrt{\Lambda/12\pi^2} \sim 10^{-33}$\,eV.
(I will use units where $\hbar=c=k=1$ unless explicitly indicated.)
Since de~Sitter space can in principle last forever, thermal fluctuations will arguably give rise to a large number of BBs.
In that case, we need to face the prospect that our leading cosmological model, carefully constructed to fit a multitude of astronomical observations and our current understanding of the laws of physics, actually predicts nonsense.

As we will see, the situation is more subtle than this discussion implies; nevertheless, the BB problem is a real one and can act as a constraint on otherwise tenable cosmological models.
The real problem is not ``this model predicts more BBs than OOs, but we are not BBs, therefore the model is ruled out."
The problem is that BB-dominated models are self-undermining, or cognitively unstable -- we cannot simultaneously believe that such a model is correct, and believe that we have good reasons for such a belief.

In this paper I will attempt to clarify a number of potentially confusing issues about the BB problem.
In the next section, we will discuss what it actually means to be a Boltzmann Brain, or another kind of random fluctuation from a high-entropy state.
Then we will confront the question of what kinds of realistic cosmological scenarios actually predict the existence of BBs, arguing that the existence of a positive cosmological constant is not sufficient.
In the subsequent two sections we will go over the argument as to why cosmologies dominated by BBs should be rejected as realistic models of the world, and address some of the counterarguments to that view.

\section{The Origin of the Problem}

To understand the Boltzmann Brain problem, it is useful to revisit the discussions between Ludwig Boltzmann and his contemporaries in the early days of statistical mechanics near the end of the nineteenth century.
Boltzmann and others showed how to recover the major results of thermodynamics by considering gases to be large numbers of randomly-moving particles.
In particular, the Second Law, which states that the entropy $S$ of a closed system will only ever stay constant or increase over time, was shown to be a matter of overwhelming probability rather than absolute certainty.

One way of understanding this result is to use the definition of entropy that is inscribed on Boltzmann's tombstone:
\be
  S = k \log W.
\ee
Here $k$ is Boltzmann's constant, and $W$ is the phase-space volume of a macrostate.
That is, we imagine taking phase space (the space of all microstates of the theory, in this case the position and momentum of every particle) and partitioning it into sets called ``macrostates.''
One common way of doing this is to define some macroscopically observable features of the system, perhaps up to some given precision, and declaring a macrostate to be the set of all microstates with the same observable properties.
Then $W$ can be thought of as the number of microstates in such a macrostate, and the entropy $S$ is the logarithm of this number.
It now makes sense why, starting from some low-entropy initial condition, the Second Law is likely to hold: there are, by construction, many more high-entropy states than low-entropy ones.
There will generally be a probability that the entropy of a system will fluctuate downward, from $S_1$ to $S_2 < S_1$, but that probability will be exponentially suppressed by $\Delta S = S_1 - S_2$:
\be
  P(\Delta S) \propto e^{-\Delta S}.
  \label{deltaS}
\ee
In realistic systems with many particles, this probability becomes negligibly small, so the Second Law is extremely reliable for all practical purposes.
(In very small systems with few moving parts, fluctuations can of course be important \cite{evanssearles,jarzynski}.)

From the very beginning of this program, objections were raised to its fundamental assumptions.
One famous objection was based on the recurrence theorem, put forward by Poincar\'e in 1890.
Starting from any initial state and evolving for a sufficiently long time, any classical Hamiltonian system with a bounded phase space will eventually return arbitrarily closely to the original state \cite{prt}.
(The quantum version of the theorem posits finite-dimensional Hilbert space rather than bounded phase space \cite{qrt1,qrt2,qrt3,Dyson:2002pf}.)
Poincar\'e himself realized the implications for statistical mechanics:
\begin{quote}I do not know if it has been remarked that the English kinetic theories can extract themselves from this contradiction.  The world, according to them, tends at first toward a state where it remains for a long time without apparent change; and this is consistent with experience; but it does not remain that way forever, if the theorem cited above is not violated; it merely stays that way for an enormously long time, a time which is longer the more numerous are the molecules.  This state will not be the final death of the universe, but a sort of slumber, from which it will awake after millions of millions of centuries.  According to this theory, to see heat pass from a cold body to a warm one, it will not be necessary to have the acute vision, the intelligence, and the dexterity of Maxwell's demon; it will suffice to have a little patience \cite{prt}.
\end{quote}
Equilibration is not forever, only for a very long time -- recurrence times are typically exponential in the maximum entropy of the system.
The recurrence theorem implies that eventually, even a system in its highest-entropy configuration will evolve back to a low-entropy initial state, in seeming violation of the Second Law.

It was Zermelo who turned Poincar\'e's mischievous remark about recurrence into a full-blown objection to Boltzmann's understanding of the Second Law \cite{zermelo1,zermelo2}.  His argument was simple:
the recurrence theorem implies that a graph of entropy vs.\ time must be a periodic function,
while the Second Law states that the entropy must monotonically increase, and both cannot
be simultaneously true.  Zermelo believed that the Second Law was absolute, not merely
statistical, and concluded that the mechanistic underpinning to thermodynamics offered
by kinetic theory could not be valid.

Boltzmann was absolutely convinced that Zermelo's objection was unfounded, but he suggested multiple ways out of the apparent contradiction, and not all of the strategies were mutually consistent \cite{boltzmannvzermelo1,boltzmannvzermelo2}.
One of his proposals was very close to our current understanding of the problem: the Second Law might not always hold in an infinitely old universe, but it can be perfectly applicable in one that has a beginning point in an initially low-entropy state:
\begin{quote}
The second law will be explained
mechanically by means of assumption $A$ (which is of course unprovable) that the universe,
considered as a mechanical system -- or at least a very large part of it which surrounds
us -- started from a very improbable state, and is still in an improbable state \cite{boltzmannvzermelo2}.
\end{quote}
The remark is suggested almost off-handedly, but it has deep implications.
The idea that the universe ``started'' in any state at all, rather than having existed forever, is not at all obvious, and indeed seems to run counter to a straightforward understanding of conventional Newtonian mechanics.
Boltzmann was writing in 1896, well before Einstein's formulation of general relativity or Lema\^itre's theory of the Primeval Atom, now known as the Big Bang.\footnote{We might wonder why neither Boltzmann nor his contemporaries pursued this implication further, suggesting some kind of Big Bang even before the discoveries of general relativity or the expansion of the universe.}
Modern treatments of the foundations of statistical mechanics generally accept the need for a ``Past Hypothesis'' according to which our universe started in a low-entropy state \cite{albert,Carroll:2010zz}.

But another suggestion of Boltzmann's went in a completely different direction.
Perhaps the universe is actually infinitely old, and on very large scales is in thermal equilibrium.
Then, just as Poincar\'e noted, rare fluctuations (in space or in time) would drive it away from a state of maximal entropy.
Boltzmann's twist was to turn lemons into lemonade: maybe we live in the aftermath of just such a fluctuation.
\begin{quote}
There must then be in the universe, which is in thermal
equilibrium as a whole and therefore dead, here and there relatively small regions
of the size of our galaxy (which we call worlds), which during the relatively short
time of eons deviate significantly from thermal equilibrium.  Among these worlds the state
probability increases as often as it decreases.  For the universe as a whole the two
directions of time are indistinguishable, just as in space there is no up or down.
However, just as at a certain place on the earth's surface we can call ``down'' the direction 
toward the centre of the earth, so a living being that finds itself in such a world at a certain
period of time can define the time direction as going from less probable to more probable
states (the former will be the ``past'' and the latter the ``future'') and by virtue of this
definition he will find that this small region, isolated from the rest of the universe, is
``initially'' always in an improbable state \cite{boltzmannvzermelo2}.
\end{quote}
(In a slightly earlier paper, he attributes this idea to ``my old assistant, Dr. Schuetz" \cite{boltzmannschuetz}.)

In this passage, Boltzmann anticipates ideas that are often thought of as contemporary and controversial.\footnote{Quite a similar cosmological model, also based on order arising out of the random motions of disordered matter, was proposed by a much earlier atomist, the Roman poet and philosopher Lucretius, in the first century BCE \cite{lucretius}.}
He imagines a vast cosmos where local conditions are very different in different regions, what a modern physicist would call ``the multiverse.''
He notes that most of this universe is ``in thermal equilibrium as a whole and therefore dead,'' and hence that it is natural we would find ourselves in a relatively atypical region that had departed from equilibrium. 
This is arguably the first application of the anthropic principle in modern science.
Finally, of course, Boltzmann has perfectly internalized the highly counter-intuitive fact that the arrow of time (a concept not fully elucidated until 1927, by Eddington) is completely tied to the thermodynamic Second Law -- we define the past as the direction of time in which entropy is lower, and the future as the direction in which it is higher.

Interestingly, Boltzmann did not seem to recognize an apparently fatal flaw in his scenario. 
That was left to Eddington, who in 1931 pointed out that anthropic reasoning only requires the existence of intelligent observers, not an entire galaxy such as the one in which we live.
According to (\ref{deltaS}), fluctuations into low-entropy states are rare, and fluctuations into even-lower-entropy states are exponentially more rare than that.
Consequently, given any particular anthropic criterion, Boltzmann's fluctuation scenario should predict that we live in a universe that has the highest possible entropy, conditioned on that criterion being met.
Eddington phrased this objection in terms of a particular anthropic criterion: the existence of mathematical physicists (presumably ones who could work out exactly this calculation):
\begin{quote}
A universe containing mathematical physicists [under these assumptions] will at any assigned date be in the state of maximum disorganization which is not inconsistent with the existence of such creatures \cite{eddington}.
\end{quote}
Such a universe would be extremely different from our current one.
Richard Feynman thought that these points -- the puzzle of the low-entropy conditions near the Big Bang, and the inadequacy of Boltzmann's fluctuation scenario at addressing the problem -- were sufficiently important that they should be familiar to first-year undergraduates at Caltech, so he makes the point in the \emph{Feynman Lectures on Physics}:
\begin{quote}
[F]rom the hypothesis that the world is a fluctuation, all of the predictions are that if we look at a part of the world we have never seen before, we will find it mixed up, and not like the piece we just looked at. If our order were due to a fluctuation, we would not expect order anywhere but where we have just noticed it.

We therefore conclude that the universe is \emph{not} a fluctuation, and that the order is a memory of conditions when things started. This is not to say that we understand the logic of it. For some reason, the universe at one time had a very low entropy for its energy content, and since then the entropy has increased. So that is the way toward the future. That is the origin of all irreversibility, that is what makes the processes of growth and decay, that makes us remember the past and not the future, remember the things which are closer to that moment in history of the universe when the order was higher than now, and why we are not able to remember things where the disorder is higher than now, which we call the future \cite{feynman}.
\end{quote}

More recently, Albrecht and Sorbo \cite{Albrecht:2004ke,Albrecht:2002uz}, drawing on a discussion by Barrow and Tipler \cite{barrowtipler}, took this reasoning to its logical conclusion.
If our anthropic criterion is simply ``the existence of intelligent observers,'' Boltzmann's fluctuation scenario predicts that the overwhelming majority of such observers should be the absolutely smallest deviations from thermal equilibrium that is compatible with one's definition of an intelligent observer.
Plausibly, such a being would require only those body parts that are absolutely essential to cognition and consciousness -- it would resemble a disembodied brain surrounded by thermal equilibrium.
These minimal observers have become known as Boltzmann Brains (BBs).

The standard concern about the Boltzmann fluctuation scenario, as reflected in the remarks by Eddington and Feynman, is that the high-entropy universe it predicts (with very large probability) looks very different from the relatively orderly cosmos in which we find ourselves. 
The BB phenomenon is meant to take this expectation to its logical extreme: not only would the universe look different, but typical observers in the universe would look utterly different from us. 
But as we shall see, it will pay to be a little more careful about the precise reasons why this kind of cosmological scenario deserves to be rejected, if indeed it does.

\section{What Are Boltzmann Brains?}

A Boltzmann Brain is a configuration of matter that is (along with its local environment) as close as possible to thermal equilibrium, while still qualifying as an intelligent observer.
One might naturally worry that reasoning quantitatively about such a concept would be nearly hopeless, given the difficulty in specifying what it means to be an ``intelligent observer."
In practice, however, such worries turn out not to be very important, due to the kinds of numbers involved.
Given any reasonable physical criteria for what constitutes a conscious brain, the ratio of BBs to ordinary observers (OOs) in a given cosmological model is almost always going to be very close to either 0 or 1.

We can see this in toy examples.
Consider two (unrealistic, but illustrative) fiducial universes, a short-lived one $A$ and a long-lived one $B$.
Imagine that both models describe a spatial volume of order that of our currently observable universe, \emph{i.e.}\ approximately the Hubble length cubed:
\be
  V \sim (H_0^{-1}c)^3 \sim 10^{78}\, \mathrm{cm}^3,
\ee
where $H_0$ is the Hubble constant and $c$ is the speed of light.
(As we will soon see, order-of-magnitude estimates are more than adequate in this game.)
{\bf Universe} $\boldsymbol A$ we imagine to only last about the current age of the universe,
\be
  t_A \sim H_0^{-1} \sim 10^{18} \, \mathrm{s}.
\ee
Even if our universe doesn't come to an end in a big crunch, its current state could plausibly change dramatically in a phase transition \cite{Page:2006dt,Page:2009mc}, which would count as ``ending'' for these purposes.
{\bf Universe} $\boldsymbol B$, meanwhile, we take to persist for the recurrence time appropriate for a system with an entropy given by the de~Sitter entropy of our universe, 
\be
  t_B \sim e^{S_{\dS}} \sim e^{10^{122}}\, \mathrm{s}.
\ee
This is an extremely long time.
So much so that it doesn't really matter what units we use to measure it with; the number is essentially the same whether measured in Planck times or Hubble times:
\be
  t_B\left(\frac{H_0^{-1}}{t_P}\right) \sim e^{10^{122}}\left(\frac{10^{18} \, \mathrm{s}}{10^{-43} \, \mathrm{s}}\right) 
  \sim e^{(10^{122} + 10^2)} \sim e^{10^{122}}.
\ee
For this reason, in these kinds of calculations it is convenient to interpret the $\sim$ symbol as meaning ``within $\pm 1$ in the top exponent."

Now let us compare the number of OOs in these universes to the number of BBs. 
We can very roughly estimate the number of OOs by taking the number of stars in our observable universe, $n_*\sim 10^{24}$, and imagining that the average number of ordinary observers coming into existence around each such star within a Hubble time is some large number $\bar{N}_\calo$.
The number of people who have lived on Earth is roughly $10^{11}$, so the minimum value of $\bar{N}_\calo$ is $10^{-13}$.
Let's posit as a reasonable maximum $\bar{N}_\calo \leq 10^{100}$; the exact number will depend on one's attitude toward conscious observers living within computer simulations constructed by advanced civilizations.
By definition, OOs live in the thermodynamically conventional aftermath of a low-entropy Big Bang, not as fluctuations at much later times, so the number of OOs in universes $A$ and $B$ is the same:
\be
  N_\OO(A) = N_\OO(B) = n_* \bar{N}_\calo,
\ee
so that
\be
  10^{11} \leq N_\OO(A,B) \leq 10^{124}.
  \label{noo}
\ee

The number of BBs, meanwhile, is given by the spacetime volume times the rate at which BBs fluctuate into existence.
The latter number will be of order $e^{-\Delta S}$, where $\Delta S$ is the difference in entropy between the equilibrium state and the state that is almost equilibrium except for a single brain.
(More than one brain could simultaneously fluctuate into existence, but that is sufficiently unlikely that we can ignore the possibility.)
For simplicity let us model the universe as a thermal gas of relativistic particles with a temperature $T_\dS \sim 10^{-33}$\,eV (so the particles would essentially be all photons and gravitons).
The entropy of a region of space filled with such a gas is of order the number of particles, and the energy per particle is of order $T_\dS$.
The entropy decrease when a BB is created is therefore simply the number of thermal particles removed from the gas in order to make it; for a BB of mass $M$ we have
\be
  \Delta S = \frac{M}{T_\dS}. 
\ee
If we imagine that a functioning brain needs to include at least Avogadro's number ($6\times 10^{23}$) protons and neutrons, this gives us
\be
  \Delta S \sim 10^{66}.
\ee
The rate of fluctuation into a BB is therefore something like
\be
  \Gamma_\BB \sim e^{-\Delta S} \sim e^{-10^{66}}.
\ee
(This is, again, a quantity for which units are irrelevant.)
We therefore have
\begin{align}
  N_\BB(A) = \Gamma_\BB t_A V \sim e^{-10^{66}} \times 10^{96} \sim 0 ,
  \label{nbba}
\end{align}
while
\begin{align}
  N_\BB(B) = \Gamma_\BB t_B V \sim e^{-10^{66}} \times e^{10^{122}} \sim e^{10^{122}}. 
  \label{nbbb}
\end{align}

We therefore obtain the fairly robust conclusion
\be
  N_\BB(A) \ll N_\OO(A,B) \ll N_BB(B).
\ee
This kind of result is common in BB calculations.
Universe $A$, which only lasts for a Hubble time, has almost no BBs, so OOs always win out; universe $B$, which lasts for a recurrence time, is overwhelmingly dominated by BBs.
For this reason, what tends to matter is not the specifics of what constitutes a BB (or how many OOs might flourish in the aftermath of the Big Bang), but rather whether BBs fluctuate into existence at all, and whether the universe lasts for a short time or essentially forever.
(For a more careful discussion of what might constitute a BB, and the rate at which they are produced, see \cite{DeSimone:2008if}.)

Here are some salient features of the process by which BBs would fluctuate into existence.
For this purpose let us imagine again that we have a thermal gas of real particles; in the next section we will ask whether that's a reasonable model of the ultimate de~Sitter vacuum.

\begin{itemize}
\item{}\emph{The raw materials for BBs would be pair-produced from photons and gravitons.}
Boltzmann imagined a gas of atoms floating around, which could randomly arrange themselves into low-entropy configurations.
In a modern cosmological scenario, the temperature of empty de~Sitter space is extremely low, much lower than the mass of the lightest massive particles.
(For current cosmological parameters, at any rate; it is also interesting to think of BB creation in other parts of the multiverse, where there might be much larger vacuum energy and correspondingly higher temperatures.)
The eventual empty universe is therefore dominated by massless particles -- photons and gravitons, neither of which are likely to assemble into conscious observers, since their interactions are so weak that it is hard to imagine forming bound states other than black holes.
Nevertheless, there can be collisions between rare high-energy photons or gravitons, which could pair-produce electrons and positrons, or protons and anti-protons, and so forth.
The general tendency of such pairs would be to re-annihilate rather quickly, but occasionally the new particles will have enough momentum to travel far apart from each other.
Sometimes (rarely, as should be henceforth understood) many such collisions will happen nearby, producing enough nearby matter to assemble itself into a macroscopic object such as a brain.

\item{}\emph{BBs are assembled slowly.}
It's tempting to think of the fluctuation process as something relatively rapid: a BB materializes into existence in a flash of colliding high-energy particles, after which it lives and thinks for however long it is able to in its rather harsh environment.
This story is unjustifiably asymmetric in time.
If the macroscopic BB configuration does not itself have a good deal of time-directedness (such as angular momentum), it's not hard to show that most fluctuation processes are statistically identical forward and backward in time around the moment of minimum entropy \cite{Aguirre:2011ac}.
The way that BBs usually get assembled, in other words, is just the time-reverse of how they will ultimately decay.
If the decay process is one of first freezing, then gradually disassembling, the creation process will involve relatively gradual assembly of individual pieces, followed by a relatively quick heating-up.
There may be interesting bio-chemical-physical questions about how long it would take a BB to completely dissipate in empty space; the point is that, however long that is, the assembly process takes just as long.
(If the BB is conscious and experiences the flow of time after it is created, it is natural to think that it was likely to have been conscious and experiencing a ``reversed'' arrow of time during the creation process, although the precise link between the thermodynamic and psychological arrows of time is controversial \cite{schulman,maroney}.)

\item{}\emph{Any specified class of macroscopic object can fluctuate into existence.}
The original impetus for Boltzmann's suggestion came from the Poincar\'e recurrence theorem, but the BB problem relies only on random fluctuations, not precise recurrence of any given configuration.
The relevant property is therefore ergodicity -- the system will explore the entirety of its phase space, given enough time -- rather than recurrence.
Not all physical systems are ergodic, but to bring the BB problem to life we only need local subsystems to behave ergodically, not the universe as a whole. 
That seems extremely reasonable (though I don't know of a theorem), given that subsystems are generally open and receive random perturbations from their surrounding environments.

\item{}\emph{The most common fluctuations will be as close to equilibrium overall as possible.}
We expect not only Boltzmann Brains, but also Boltzmann People, Boltzmann Solar Systems, Boltzmann Galaxies, or even Boltzmann Universes, to eventually fluctuate into existence.
Given any particular criterion that we may look for in a fluctuation (such as ``the existence of an intelligent observer,'' or ``the existence of a civilization that lasts for at least a million years"), the probability formula (\ref{deltaS}) implies that the most common such fluctuation will be the one with the highest entropy overall.
In ordinary, thermodynamically-sensible evolution from low-entropy initial conditions, real physical structures often 
contain substantial mutual information -- there are strong correlations between, for example, the state of a camera that has just taken a picture and the state of whatever it was pointed at when the image was taken.
There is no reason to expect such correlations in Boltzmann fluctuations, if they are not specifically sought after as part of the criteria defining the fluctuations we are considering.
If a camera with an image in its memory (or a person with a memory in their brain) fluctuates into existence, it is overwhelmingly likely that the rest of the universe will still be in thermal equilibrium.

One might imagine raising a ``Darwinian'' objection to the BB problem: ``the most likely way that a complex biological organ such as a brain can appear is not through random fluctuations, but via biological evolution under the pressure of natural selection."
This objection misunderstands the role of entropy in the situation.
The environment of our biosphere, in which actual brains evolve, is an open system driven very far from equilibrium, as a direct result of its descent from the extremely low-entropy early universe.
In that context, biological evolution is naturally the best way of creating brains.
But starting from equilibrium, creating brains by first fluctuating into the state of our early universe (or even just of our biosphere) is enormously less likely than creating individual brains uncorrelated with their environments.

\item{}\emph{The BB scenario relies on the fact that the laws of physics are Markovian (memoryless).}
As with correlations in space at one moment, ordinary thermodynamically-sensible evolution also induces correlations across time.
A photograph of a birthday party from twenty years ago is highly correlated with what happened at the party.
Boltzmann fluctuations have no such correlations; just because a photo fluctuates into existence, it doesn't follow that a birthday party must have fluctuated years earlier.
We can say this with confidence because the current state of a physical system suffices, according to our best known laws of physics, to determine its subsequent evolution.
In the context of stochastic processes this feature is known as the ``Markov property,'' but ever since Laplace we have appreciated that the laws of physics seem to be memoryless in this sense.
Therefore, the existence of a memory in a BB at the present time -- including the expectation that the BB will act as if the memory is real and reliable, to the best of its knowledge -- is a statement about the current state of the BB, without any strong connection to past events. 
There is no reason for a BB's impressions of either the past, or the current state of the world elsewhere, to be at all accurate.
\end{itemize}

\section{What Theories Predict Boltzmann Brains?}

Boltzmann envisioned an eternal Newtonian universe, with a fixed background spacetime and an infinite number of randomly-fluctuating particles.
The shift to a general-relativistic perspective, where spacetime is dynamical and responds to energy and matter, changes the picture dramatically.
Our universe is expanding from a hot Big Bang state of about 14 billion years ago.
Classically, there are two natural possibilities.
The universe could eventually recollapse, in which case its lifetime is very likely to be much less than the time for even a single BB to fluctuate into existence, and there is not a BB problem.
Alternatively, the universe could expand forever.
Then we expect the density of matter to dilute away to zero, leaving nothing around from which BBs could be created.
Either way, for most of the 20th century cosmologists were not worrying about Boltzmann Brains.

In 1998 we found strong evidence \cite{Riess:1998cb,Perlmutter:1998np}, since verified by a number of independent techniques, that our universe is accelerating.
The simplest and most likely (though not only) explanation for this is the existence of a constant vacuum energy, equivalent to Einstein's cosmological constant $\Lambda$.
The cosmological constant is plausibly (though, again, not necessarily) constant for all time.
In that case, we would seem to have resolved the choice between recollapsing and ever-expanding universes, in favor of perpetual expansion.
The cosmic no-hair theorem \cite{Wald:1983ky} implies that our universe will evolve toward the geometry of de~Sitter space, the maximally symmetric solution to general relativity with a positive cosmological constant.

By itself that doesn't change the status of the BB problem -- classically, space still empties of particles, even in the presence of vacuum energy.
But quantum mechanics now enters in a significant way.
Like a black hole, de~Sitter space has a horizon; unlike a black hole, we think of ourselves as being surrounded by the horizon, rather than looking at it from outside.
Objects that are sufficiently far away (roughly the Hubble radius, $H_0^{-1} \sim 10^{10}$ light-years) cannot send signals to us even at the speed of light, since the space in between is expanding too rapidly.
In the 1970's, Hawking showed that quantum effects near black hole horizons cause black holes to give off black-body radiation with a temperature inversely proportional to the hole's mass \cite{Hawking:1976de}.
Soon thereafter, Gibbons and Hawking \cite{Gibbons:1977mu} showed that the de~Sitter horizon acts in a similar way, producing thermal radiation with a temperature $T_\dS = \sqrt{\Lambda/12\pi^2}$.
Put somewhat more carefully, a particle detector in its ready state will, in the quantum vacuum of de~Sitter, detect particles with such a spectrum.

This creates a somewhat surprising situation.
While classically a universe dominated by a positive cosmological constant simply empties out and evolves to zero temperature, quantum-mechanically it asymptotes to a fixed nonzero temperature.
Such a universe resembles quite closely Boltzmann's original idea: an eternal thermal system with statistical fluctuations.
It is therefore reasonable to worry that BBs will be produced in the eventual future, and dominate the number of intelligent observers in the universe.
Note that this conclusion doesn't involve speculative ideas such as eternal inflation, the cosmological multiverse, or the string theory landscape -- it refers to ordinary $\Lambda$CDM, the best-fit model constructed by cosmologists to describe the universe we live in today.
We therefore face the prospect that our best modern cosmological model is internally incoherent.
(The BB problem is also, needless to say, an important criterion for judging the empirical viability of different manifestations of the multiverse.
In that context it is often the case that the question depends sensitively on one's favorite approach to the ``cosmological measure problem'' -- how to compare the relative number of observers of different types in a large multiverse \cite{Bousso:2006ev,Bousso:2006xc,Linde:2006nw,Vilenkin:2006qg,Garriga:2007wz,Bousso:2007nd,DeSimone:2008if,Freivogel:2011eg,Salem:2011qz,Vilenkin:2011yx}.
That issue is outside the scope of our discussion here.)

This conclusion is, at least plausibly, too quick, as argued by Boddy, Carroll, and Pollack (BCP) \cite{Boddy:2014eba} (see also Gott \cite{Gott:2008ii}).
The quantum vacuum state in a de~Sitter spacetime background is a \emph{stationary} state, one that doesn't evolve with time.
(In conventional cosmological coordinate systems space expands in de~Sitter, but there are coordinates in which space is static, reflecting the existence of a timelike Killing vector.)
If we take a horizon-sized patch of space and trace out degrees of freedom outside, the density matrix describing the patch is thermal, as expected from the Gibbons-Hawking calculation.
But that thermal density matrix is also stationary; there are no time-dependent fluctuations.
BCP therefore argued that there aren't true dynamical processes in the de~Sitter quantum vacuum state that would produce BBs, so the problem simply doesn't arise.

There is a possibility of confusion here, due to the sloppy way in which physicists use the word ``quantum fluctuation."
It's helpful to distinguish between three different notions \cite{Boddy:2015fqa}:
\begin{itemize}
\item{\bf Vacuum fluctuations} are the differences between a quantum state and some appropriate classical analogue. These are captured by the notion of a non-zero variance for observable properties; given a state $|\psi\rangle$ and some observable $\widehat\calo$, we have $\langle\widehat\calo^2\rangle_\psi\geq \langle\widehat\calo\rangle^2_\psi$. The position of an electron in an atomic orbital, for example, is not localized to a position eigenstate. Vacuum fluctuations are present even in stationary states, but do not represent time-dependent dynamical processes. Examples include the Casimir effect and radiative corrections to physical parameters in quantum field theory.
\item{\bf Measurement-induced fluctuations} are the stochastic observational outcomes resulting from repeating the process of putting a system into some state, measuring some property of it, then returning it to the original state. 
\item{\bf Boltzmann fluctuations} are true dynamical processes in which entropy decreases, resulting from stochastic dynamics in time-dependent states. They can be present if a microstate is time-dependent even if its corresponding coarse-grained macrostate is not; {\it e.g.} a box of gas in equilibrium can fluctuate to a lower-entropy state. BBs would be examples of Boltzmann fluctuations.
\end{itemize}
BCP argue that there are vacuum fluctuations in de~Sitter space, and there would be measurement-induced fluctuations in the presence of an out-of-equilibrium measuring device; but there is no such measuring device in the vacuum, and there are no Boltzmann fluctuations in a stationary state.
Therefore, BBs don't appear in the true de~Sitter vacuum.
This argument was made in the context of Everettian (many-worlds) quantum mechanics, but similar arguments apply to Bohmian mechanics \cite{Goldstein:2015mha}.

This conclusion is not uncontroversial.
It relies on a particular construal of what ``happens'' in a stationary quantum state: {\it i.e.}, that nothing really happens at all, in contrast to conventional quantum measurement processes, which involve decoherence, branching of the wave function, and an increase of entropy in the system's density matrix.
An alternative perspective would be to argue that something ``happens'' if it occurs within a history that is part of a decoherent set of histories.
As Lloyd has argued \cite{Lloyd:2016ahu}, it is straightforward to construct sets of decoherent histories in which all sorts of interesting dynamical processes take place, even if the overall density matrix corresponds to a pure stationary quantum state.
Taking this as our criteria for the existence of BBs, it follows that BBs are inevitable in almost any cosmological model.
Indeed, BBs would occur even in the Minkowski vacuum, with zero cosmological constant and zero temperature; this possibility has been considered \cite{Page:2005ur,Page:2006dt,Davenport:2010jy,Nomura:2015zda}, but it seems very different from the standard picture of thermal de~Sitter fluctuations.

Relatedly, the absence of dynamical observers in a stationary state also relies on the global branching structure of the Everettian wave function.
Given any density matrix for a subsystem, there are a large number of different possible purifications in the larger Hilbert space.
Bennett and Riedel \cite{bennettriedel} have pointed out that one could imagine purifications that entangled truly dynamical states of any given de~Sitter patch (including BBs or OOs of whatever kind) with decohered pointer states in the external environment, while maintaining the feature that the full de~Sitter patch density matrix was stationary and thermal.
In that case a conscientious Everettian would judge that there really were dynamical observers within the patch, even though their existence could not be discerned by studying the density matrix alone.
It seems unlikely that this kind of state is actually what arises in conventional cosmological evolution, but both of these potential loopholes in the claim that there are no BBs in the de~Sitter vacuum are worthy of further study.

The idea that BBs are absent in the stationary de~Sitter vacuum is only relevant if cosmological evolution brings us to that vacuum state.
In the context of quantum field theory in curved spacetime, this is what we expect to happen, at least in local regions of space, according to quantum versions of the cosmic no-hair theorem \cite{Marolf:2010nz,Hollands:2010pr}.
It's possible, however, that the situation changes dramatically in true quantum gravity, as argued by Dyson, Kleban, and Susskind (DKS) \cite{Dyson:2002pf}.

Just as with a black-hole horizon, the de~Sitter horizon can be associated with an entropy proportional to its area:
\be
  S_\dS = \frac{A}{4G} = \frac{3\pi}{G\Lambda}.
\ee
For the measured vacuum energy, this works out to approximately $S_\dS \sim 10^{122}$.
According to the philosophy of horizon complementarity \cite{Susskind:1993if}, the interior of the horizon is therefore a system described by a finite-dimensional Hilbert space, with
\be
  \mathrm{dim}\,{\calh} \approx e^{S_\dS} = e^{10^{122}}.
\ee
(See also \cite{Banks:2002wr,Albrecht:2004ke,Parikh:2004wh,Banks:2006rx}.)
By itself this doesn't bear directly on the question of BBs; a finite-dimensional open system can asymptote to its vacuum state by dissipating excitations into its surrounding environment.
But horizon complementarity suggests a more radical viewpoint, where this finite-dimensional system should be thought of as describing the entire theory; everything we would conventionally think of as ``the rest of the universe'' is encoded as information on the horizon itself.
Different observers will describe the location of the de~Sitter horizon differently, but all will give compatible descriptions of regions of spacetime that overlap between different horizons.

Finite-dimensional Hilbert spaces are the quantum analogue of bounded phase spaces in classical mechanics, and they have a corresponding recurrence theorem.
If evolution is unitary and Hilbert space is finite-dimensional, the state will not simply evolve to the de~Sitter vacuum and stay there, as it would if Hilbert space were infinite-dimensional.
Observable deviations from the vacuum might be extremely small, but they would always be present.
Therefore, we expect not only recurrence, but Boltzmann fluctuations into all manner of low-entropy states.
DKS pointed out that the presence of such fluctuations renders $\Lambda$CDM unacceptable as a realistic cosmological model (if we accept the appropriate interpretation of horizon complementarity), and Albrecht and Sorbo \cite{Albrecht:2004ke} connected this directly to the Boltzmann Brain problem.

One loophole in this analysis appears if our current vacuum state with its cosmological constant is only one of many vacua in the theory, including a Minkowski vacuum with $\Lambda = 0$.
That is quite plausible in a number of theories, including the landscape of string theory, where different ways that extra dimensions can be compactified lead to a proliferation of metastable vacuum states \cite{Bousso:2000xa,Kachru:2003aw,Susskind:2003kw}.
As long as there is at least one $\Lambda=0$ vacuum, the Hilbert space of the full theory will be infinite-dimensional.
It is then conceivable, as suggested by BCP, that the dynamics in a de~Sitter vacuum are effectively dissipative, and the state will settle into an asymptotically stationary quantum state, free of BBs. 

There is also a more direct way in which the existence of other vacua could save us from the BB problem: if the decay rate of a de~Sitter vacuum into a lower-energy one is sufficiently fast that BBs don't have time to form \cite{Page:2006dt,Page:2009mc}.
Such decays happen via bubble nucleation: a microscopically small region of space undergoes a quantum tunneling to a lower-energy state, creating a spherical bubble that expands at nearly the speed of light.
If bubbles ultimately collide and fill space, the phase transition is said to have ``percolated.''
As in the examples of universes $A$ and $B$ above, a relatively short-lived universe will produce essentially no BBs, even if it is possible for them to fluctuate into existence; indeed, vacuum decay is the most straightforward way to obtain a short-lived universe such as $A$.
(The universe as a whole might last for a long time, but all that matters for the BB problem is the amount of time it spends in the appropriate de~Sitter vacuum.)

Percolation is harder to achieve than one might think, for the simple reason that the space in between the bubbles is expanding and accelerating.
Roughly speaking, the bubble nucleation rate must be of order one per Hubble spacetime volume or greater.
Otherwise, even as new bubbles come into existence, the spatial volume in the original vacuum continues to grow exponentially, and the transition never completes, leaving room for an unlimited number of future BBs.
This raises the somewhat alarming prospect that the vacuum in which we currently find ourselves might be unstable over timescales comparable to the current age of the universe.
Interestingly, the measured properties of the Higgs boson suggest that its potential is conceivably unstable over just such timescales \cite{Espinosa:2007qp,ArkaniHamed:2008ym,Alekhin:2012py,Buttazzo:2013uya,Boddy:2013qma}.

The bad news is that life as we know it would be eliminated by decay to a lower-energy vacuum, which would leave behind nothing but empty space.
As Coleman and de~Luccia memorably put it in their analysis of false-vacuum decay,
\begin{quotation}
The possibility that we are living in a false vacuum has never been a cheering one to contemplate. Vacuum decay is the ultimate ecological catastrophe; in the new vacuum there are new constants of nature; after vacuum decay, not only is life as we know it impossible, so is chemistry as we know it. However, one could always draw stoic comfort from the possibility that perhaps in the course of time the new vacuum would sustain, if not life as we know it, at least some structures capable of knowing joy. This possibility has now been eliminated \cite{Coleman:1980aw}.
\end{quotation}
Fortunately, even if the nucleation rate is as high as 1 per current Hubble volume per Hubble time, there is only a $10^{-8}$ chance that a bubble of lower-energy vacuum will intersect the Earth over a century-long human lifetime.

\section{Why Are Boltzmann Brains Bad?}

The standard (but not quite correct) argument that cosmologies dominated by BBs are unacceptable is fairly straightforward: in such a universe, I would probably be a Boltzmann Brain, and I'm not, therefore that's not the universe in which we live.

Many people fail to find the standard argument convincing, on the grounds that the existence of observers elsewhere in the universe and very different from themselves shouldn't call into question the success of a model that seems to correctly account for what they see around them (see \emph{e.g.} \cite{Banks:2002wr,Gott:2008ii}).
Consider a conversation between a scientist \emph{W} who worries that BBs render otherwise reasonable cosmological scenarios unacceptable, and another scientist \emph{S} who is more sanguine.
\begin{quotation}
\emph{W}: I am worried that our best cosmological model predicts that typical observers are Boltzmann Brains.

\emph{S}: Why? Sure, there may be a huge number of BBs in this universe, but I'm not one of them, so what do I care?

\emph{W}: How do you know you're not a Boltzmann Brain?

\emph{S}: I don't seem very brain-like. I have arms, legs, etc. The environment around me seems pretty dramatically far from equilibrium.

\emph{W}: Fine. But in a large, randomly-fluctuating universe, even if you are not a minimal-fluctuation observer (a BB), it's overwhelmingly likely that you are in a state that represents the minimum possible fluctuation away from equilibrium, \emph{conditioned on whatever features your local environment has}. Call these ``Boltzmann Observers" (BOs). Even granting that you have arms and legs and are sitting in an office, it's still overwhelmingly likely that all of that has randomly fluctuated into existence out of equilibrium. How do you know you're not a Boltzmann Observer?

\emph{S}: Because I can see that I'm not! I can go out and, for example, observe the cosmic microwave background, which is evidence that the universe was in a much lower-entropy state several billion years ago, which isn't what I would expect if I were a BO.

\emph{W}: But even if you are observing the microwave background, you're not necessarily seeing the leftover photons from the Big Bang. You're seeing a radiation field entering your telescope right here in your nearby environment. In a fluctuating-universe scenario, it's certainly possible to see such photons, but it's overwhelmingly likely that they randomly fluctuated into existence, without any connection to an earlier low-entropy state.

\emph{S}: Well, that scenario makes a very reliable prediction, as Feynman basically pointed out: all I have to do is look in my radio telescope again, and see whether or not the background radiation is still there. If what I observed already is just a random fluctuation, there's no reason for it to persist over time; in fact it's incredibly unlikely. So let me wait a second before checking again and... Nope. There's the microwave background. I still seem to be in a thermodynamically sensible environment. I am not a BO.

\emph{W}: You don't have the right to conclude that. In a randomly-fluctuating universe, just as almost all observers are minimal-fluctuation BBs, and almost all observers with arms and legs and microwave backgrounds are minimal-fluctuation BOs subject to those macroscopic constraints, it's also the case that almost all of the observers with arms etc.\ who believe they have just waited a few seconds and failed to observe any evidence of the surrounding equilibrium are \emph{also} random fluctuations with just those properties. 
You can conditionalize on any macroscopic information you like: in a randomly-fluctuating universe, it remains overwhelmingly likely that you are at a local minimum of entropy that evolved by chance out of equilibrium.

\emph{S}: But everything I know and feel and think about the world is what I would expect if I were an ordinary observer who has arisen in the aftermath of a low-entropy Big Bang, and nothing that I perceive is what I would expect if I were a random fluctuation.

\emph{W}: And in a randomly-fluctuating universe, the overwhelming majority of people who would say exactly that are, as a matter of fact, random fluctuations. That's why those kinds of cosmological models are bad.
\end{quotation}
I will argue that cosmologies dominated by BBs should be rejected, not because I have empirical evidence that I am not one and I should be, but because such models are cognitively unstable.

\subsection{The standard argument}

Let's rehearse the standard argument in somewhat more formal terms.
Bayes's Theorem relates $P(T_i|D)$, the posterior credences we attach to a set of mutually exclusive and exhaustive propositions (``theories'') $T_i$ after gathering some data $D$, to the prior credences $P(T_i)$ and the likelihoods $P(D|T_i)$ that such data would be obtained if each proposition were true:
\be
  P(T_i|D) = \frac{P(D|T_i)P(T_i)}{P(D)},
  \label{bayes}
\ee
where the total probability of obtaining the data is $P(D) = \sum_i P(D|T_i)$.
We would like to use this expression to judge the relative credence we should assign to different cosmological models, where the ``data'' is taken to be our experience -- that we seem to be ordinary observers rather than BBs, or at least that we seem to be embedded in a certain kind of environment that is highly out-of-equilibrium and nicely consistent with thermodynamically conventional evolution from a low-entropy beginning.

Consider our two model universes, $A$ with very few BBs, and $B$ in which BBs far outnumber OOs, so that our two propositions are $T_A$ and $T_B$ (we live in universe $A$ or $B$, respectively). 
The standard argument assumes that the likelihoods for our observed data in each theory are roughly proportional to the fraction of observers that are OOs:
\be
  P(D|T_i) \sim \frac{N_\OO}{N_\OO+N_\BB(T_i)} \sim 
  \begin{cases}
  1& \quad A, \\ 
  e^{-10^{122}}& \quad B,
  \end{cases}
  \label{likelihoods}
\ee
where the numbers are taken from (\ref{noo}), (\ref{nbba}), and (\ref{nbbb}), and we are assuming comparable numbers of ordinary observers in the two theories.
This form for the likelihoods is equivalent to an assumption of ``the Copernican principle" or ``mediocrity'' or ``typicality'' or ``self-sampling'' or ``indifference'' concerning who we are in the universe: given some reference class of intelligent observers, we are equally likely to have been any of them, and should reason accordingly \cite{gott93,Vilenkin:1994ua,bostrom,Olum:2000zw}.
The numbers in (\ref{likelihoods}) are so wildly disparate that, for a very wide range of possible priors, we can immediately conclude to a high degree of confidence that we live in cosmology $A$, the short-lived one without BBs.
This formalizes the intuition that BBs must be sub-dominant because we are not ourselves BBs.

It's always possible, of course, to choose priors that can overcome even dramatic differences in likelihoods.
It might seem fair to assign priors $P(T_A)$ and $P(T_B)$ that are roughly equal to each other, on the theory that we are comparing two cosmological models of comparable plausibility.
But one might alternatively adopt a ``self-indication assumption'' (SIA), according to which we should reason as if we were chosen randomly from the set of all \emph{possible} (rather than actual) observers \cite{bostrom,Olum:2000zw}. 
Effectively, SIA can be thought of as boosting the prior for theories with BBs by an enormous amount, simply because there are a lot of observers in them:
\be
  P(T_i) \propto N_\OO + N_\BB(T_i).
\ee
This factor would then cancel the suppression in (\ref{likelihoods}), undermining the standard argument against BB-dominated cosmologies.

Bostrom \cite{bostrom} has given an argument against giving higher priors to theories with more observers in them, which he labels the ``Presumptuous Philosopher problem." 
Consider two competing theories that don't have any BBs at all, but that differ substantially in the number of predicted OOs (and are otherwise indistinguishable).
Weighting the priors by the number of observers would allow us to essentially rule out -- perhaps to extremely high confidence -- the universe with fewer observers, even without taking any data or performing any other traditional scientific investigation.
The point is that the seemingly innocent and humble assumption that ``we are typical observers'' isn't innocent at all -- it actually grants us enormous leverage in deciding between competing models, by implying that ``typical observers are like us." 
It seems presumptuous to be able to convincingly rule out simple, plausible cosmological scenarios by such non-empirical arguments, just by thinking rather than by collecting data.
Let us therefore proceed under the assumption that that the priors for theories $A$ and $B$ are comparable.

\subsection{Typicality and likelihoods}

The idea that we should reason as if we are typical observers, and therefore that the relevant likelihoods should obey (\ref{likelihoods}), has been criticized by Hartle and Srednicki (HS) \cite{Hartle:2007zv,Srednicki:2009vb} (also \cite{Azhar:2016mpg}, or for an opposing view \cite{Bousso:2007nd}).
Their argument is closely related to the Presumptuous Philosopher problem.
Imagine we learned enough about biochemistry that we were convinced the probability of life coming into existence in the atmosphere of Jupiter was 0.5, and furthermore that if such life did arise, with probability 1 it would evolve an intelligent species (perhaps the floating gas bags envisioned by Carl Sagan) with $10^{12}$ individuals alive today.
There are therefore two scenarios in front of us, with equal priors: one in which humans are the only intelligent observers in the Solar System, the other of which over 99\% of the intelligent observers are Jovians.
HS point out that adopting the likelihoods (\ref{likelihoods}), with the role of OOs and BBs replaced by humans and Jovians, respectively, would lead us to conclude that there was less than a one percent chance that the Jovians existed, even though we haven't actually gone to look for them.
That's because, in the scenario where there are any Jovians at all, a typical intelligent observer in the Solar System is a Jovian, so the fact that we are human counts as strong evidence against their existence.
Once again, we seem to have helped ourselves to very strong conclusions about the universe without looking at it.

To protect against such presumptuousness, HS argue that a theory of cosmology should be thought of as coming in two parts: a physical model of the universe $T_i$, which specifies (among other things) the number of any particular kind of observer, and a \emph{xerographic distribution} $\xi^{(i)}_a$, which for each physical model $T_i$ specifies the probability distribution for us to actually be any of the observers in the model. 
One example of a xerographic distribution would simply be uniform: among all observers, we have an equal probability to be any one of them,
\be
  \xi_\mathrm{U} = \frac{1}{N_\OO + N_\BB}.
\ee
Here we have classified every observer as either an OO -- defined as observers arising from thermodynamically sensible evolution from very low-entropy conditions, with the ability to draw more or less accurate conclusions about what those conditions were by observing features of their present environment -- or a BB (even if the observer is much more than just a brain).
But we can also consider distributions that are concentrated entirely on OOs, even in universes that have huge number of BBs in them :
\be
  \xi_\mathrm{OO} = 
    \begin{cases}
  0& \quad \mathrm{BBs}, \\ 
  \frac{1}{N_\OO} &\quad \mathrm{OOs}.
  \end{cases}
  \label{xerographic}
\ee

Therefore, we can imagine a cosmological model such as our universe $B$ (which lasts for a recurrence time), but which comes equipped with either a uniform xerographic distribution $\xi_\mathrm{U}$ or one $\xi_\mathrm{O}$ that is concentrated on OOs.
In that case, the likelihood (\ref{likelihoods}) corresponding to $(B, \xi_\mathrm{U})$ would be small as before, but the one for $(B, \xi_\mathrm{OO})$ would be of order unity.
The data that we do not seem to be BBs would count against $B$ with the uniform distribution, but $B$ with a concentrated distribution would be perfectly acceptable, and we would have no right to reject a long-lived cosmological model with dynamical fluctuations (unless the prior were chosen to be very small).

As mentioned in the above dialogue, however, we don't quite have data that we are not BBs, or at least not that we aren't Boltzmann Observers who randomly fluctuated into existence in the conditions we seem to find ourselves in.
Given only that we are some kind of observer, and a cosmology dominated by random fluctuations, it would indeed be unlikely that we would seem to find ourselves surrounded by a thermodynamically sensible out-of-equilibrium environment; but given  that we are exactly the kind of observers we are, in a randomly-fluctuating universe it is still overwhelmingly likely that our recent past isn't nearly as low-entropy as we think it is.

A useful approach to the typicality issue that avoids the Presumptuous Philosopher problem, but doesn't sweep the BB problem under the rug, is the suggestion by Neal that we apply ``Full Non-indexical Conditioning'' (FNC) to our place in the universe \cite{Neal:2006py}.
FNC can be summarized by allowing us to use all of the data we currently possess -- our personal memories (which might not be accurate, but we have them), the image we perceive of our immediate environments, etc. -- but if there are multiple observers within the resulting class, we should reason as if we are chosen from a uniform distribution within that set.
Denoting the data available to some particular observer as $D_*$, this corresponds to choosing a xerographic distribution of the form
\be
  \xi_\mathrm{FNC} = 
    \begin{cases}
  0& \quad D\neq D_*, \\ 
  \frac{1}{N_{D_*}} &\quad D = D_*.
  \end{cases}
  \label{fnc}
\ee
Within this scheme, our existence doesn't imply anything about the possibility of many intelligent observers floating in the atmosphere of Jupiter, since our data tells us we are not Jovians.

While there is a superficial resemblance between the distributions (\ref{xerographic}) and (\ref{fnc}), however, their implications are very different. 
In a cosmological model of a long-lived universe with random fluctuations, there is unlikely to be more than one OO who has exactly my data (at least within any given Hubble volume), but vast numbers of observers with that data will come from Boltzmann fluctuations.
So while conditioning on all the data I have protects me from drawing presumptuous conclusions about life on other planets, it keeps the standard Boltzmann Brain problem completely intact.

The question, then, comes down completely to what kind of priors we should attach to xerographic distributions such as (\ref{xerographic}) and (\ref{fnc}).
There is a crucial philosophical difference between them: the FNC choice (\ref{fnc}) restricts to a class of observers on the basis of the data they have, while the distribution concentrated on OOs (\ref{xerographic}) references facts that the observers themselves cannot possibly access in their current situation -- namely, whether the superficial evidence for a low-entropy past is accurate, or merely a random fluctuation.

That seems like cheating -- biasing a xerographic distribution toward a particular answer (we are OOs), not because we have any evidence for it, but because it's the answer we prefer.
After all, in a long-lived, randomly-fluctuating universe, there would be plenty of observers who arose as random fluctuations rather than from an ultra-low-entropy Big Bang, but who nevertheless drew more or less accurate conclusions about the laws of physics governing their world (even if their number would be dwarfed by the number of true Boltzmann Brains).
This would include the kinds of observers Boltzmann himself had in mind, who lived in the aftermath of a fluctuation sufficiently large to make a galaxy like ours, but not necessarily the trillions of galaxies within our observable universe.
There would even be many observers who were temporarily fooled, but then successfully updated their models when new data came in (\emph{e.g.} when the cosmic microwave background suddenly disappeared from their telescopes).
Why should we choose a large prior for a xerographic distribution that constitutes an \emph{a priori} denial that we could possibly be such observers?
It seems like we have once again donned the robes of the Presumptuous Philosopher, albeit with a different choice of fabric.

\subsection{Cognitive instability}

The standard argument against BB-dominated cosmologies is that, in such a cosmology, I should be a BB, but I'm not, therefore that's not the universe I live in.
But the analysis above undermines this reasoning somewhat.
The caution against presumptuously philosophizing suggests that we should not think of ourselves as being chosen randomly from the set of all intelligent observers in the universe; this gives us too much leverage over what the universe is like, without us actually looking at it.
But it also suggests that it's correct to think of ourselves as being chosen randomly from a very restricted class of observers -- perhaps as restricted as ``observers with precisely my macroscopic data."
This prevents us from drawing unwarranted conclusions about other observers in the cosmos, while not sneaking in any unjustified features of our own place within it.

The xerographic distribution (\ref{fnc}) we are left with, however, doesn't quite do the job of supporting the standard BB argument.
While we know (or at least seem to perceive) that we are not disembodied brains floating in an otherwise empty void, we don't know that we and our local environments haven't fluctuated into existence from a higher-entropy equilibrium state -- in a long-lived, randomly-fluctuating universe, it's overwhelmingly likely that we have indeed done so.
We can't claim to have empirical evidence against this disturbing possibility.

What we can do, however, is recognize that it's no way to go through life. 
The data that an observer just like us has access to includes not only our physical environment, but all of the (purported) memories and knowledge in our brains.
In a randomly-fluctuating scenario, there's no reason for this ``knowledge'' to have any correlation whatsoever with the world outside our immediate sensory reach.
In particular, it's overwhelmingly likely that everything we think we know about the laws of physics, and the cosmological model we have constructed that predicts we are likely to be random fluctuations, has randomly fluctuated into our heads.
There is certainly no reason to trust that our knowledge is accurate, or that we have correctly deduced the predictions of this cosmological model.

A randomly-fluctuating Boltzmann universe puts us, then, in a strange predicament.
On the one hand, we use our reasoning skills and knowledge of physics to deduce that in such a cosmos we are probably randomly-fluctuated observers, even after conditioning on our local data.
On the other hand, we can also deduce that we then have no reason to trust those reasoning skills or that knowledge of physics.

The randomly-fluctuating universe scenario is therefore self-undermining, or as Albert has characterized similar situations in statistical mechanics, \emph{cognitively unstable} \cite{albert,Carroll:2010zz,myrvold}.
If you reason yourself into believing that you live in such a universe, you have to conclude that you have no justification for accepting your own reasoning.
You cannot simultaneously conclude that you live in a randomly-fluctuating universe and believe that you have good reason for concluding that.
In (what most of us believe to be) the real world, in which there really was a low-entropy Big Bang in the relatively recent past, our memories and deductions about the past rely on a low-entropy Past Hypothesis for their justification.
In a universe dominated by Boltzmann fluctuations, such an hypothesis is lacking, and we can't trust anything we think we know.

How are we to treat the prospect of a cognitively unstable theory?
We can't say for sure that it is not true -- it's \emph{possible} that we do live in a fluctuation away from equilibrium, perhaps with completely false memories of the past.
We can't even collect evidence that would noticeably decrease our credence in the possibility, as such evidence is also likely to have fluctuated into our hands.

The best we can do is to decline to entertain the possibility that the universe is described by a cognitively unstable theory, by setting our prior for such a possibility to zero (or at least very close to it). 
That is what priors are all about: setting credences for models on the basis of how simple and reasonable they seem to be before we have collected any relevant data.
It seems unreasonable to grant substantial credence to the prospect that we have no right to be granting substantial credence to anything.
If we discover that a certain otherwise innocuous cosmological model doesn't allow us to have a reasonable degree of confidence in science and the empirical method, it makes sense to reject that model, if only on pragmatic grounds. 
This includes theories in which the universe is dominated by Boltzmann Brains and other random fluctuations.
It's not that we've gathered evidence against such theories by noticing that we are not BBs; it's that we should discard such theories from consideration even before we've looked.

It's important that we're not merely discarding the possibility that we ourselves are BBs; we are discarding the cosmological models in which they dominate.
We might ask whether this kind of reasoning should simply lead us to reject xerographic distributions that allow us to be observers whose memories are not accurate -- the universe is full of BBs, but we're not among them.
But as discussed above, this represents a criterion based on facts about our location in the universe that we ourselves do not have access to, which raises immediate problems.
As a matter of empirical observation, there are many people around us who have wildly inaccurate memories or understandings of the laws of physics; it would seem dubious to exclude them from our xerographic distribution.
In a long-lived, randomly-fluctuating universe, the dividing line between observers that have been fooled by statistically unlikely data and those than have simply drawn incorrect conclusions from the data they've collected is a hard one to draw.
And many observers with data just like ours may have been fooled thus far, but will eventually figure things out when they realize that the microwave background has disappeared.
We are on much safer footing when we choose xerographic distributions that discriminate on the basis of the currently observed situations that observers are in, rather than the accuracy of their deductions or the sensibility of their thermodynamic evolutions.

\section{Conclusions}

We therefore conclude that the right strategy is to reject cosmological models that would be dominated by Boltzmann Brains (or at least Boltzmann Observers among those who have data just like ours), not because we have empirical evidence against them, but because they are cognitively unstable and therefore self-undermining and unworthy of serious consideration.
If we construct a model such as $\Lambda$CDM or a particular instantiation of the inflationary multiverse that seems to lead us into such a situation, our job as cosmologists is to modify it until this problem is solved, or search for a better theory.
This is very useful guidance when it comes to the difficult task of building theories that describe the universe as a whole.

Fortunately, the criterion that random fluctuations dominate the fraction of observers in a given cosmological model might not be as difficult to evade as might be naively expected, if Hilbert space is infinite-dimensional and local de~Sitter phases settle into a truly stationary vacuum state, free of dynamical Boltzmann fluctuations.
That conclusion depends sensitively on how one interprets what happens inside the quantum state, an issue that is unfortunately murky in the current state of the art.
If any were needed, this gives further impetus to the difficult task of reconciling the foundations of quantum mechanics and cosmology.

\section*{Acknowledgements}

I have benefited from discussions with a large number of people over the years, including David Albert, Andy Albrecht, Anthony Aguirre, Charlie Bennett, Kim Boddy, Raphael Bousso, Jennifer Chen, Alan Guth, Jim Hartle, Matt Johnson, Andrei Linde, Don Page, Jason Pollack, Jess Riedel, and Mark Srednicki.  
This research is funded in part by the Walter Burke Institute for Theoretical Physics at Caltech, by DOE grant DE-SC0011632, by the Foundational Questions Institute, and by the Gordon and Betty Moore Foundation through Grant 776 to the Caltech Moore Center for Theoretical Cosmology and Physics.

\end{document}